\documentclass[12pt]{article}
\usepackage{epsfig}
\usepackage{a4}
 
\date{\today}

\usepackage{amsmath}

\usepackage{amsmath} 

\begin{document}

\date{}
\newcommand{\dd}{\mbox{d}}
\newcommand{\tr}{\mbox{tr}}
\newcommand{\la}{\lambda}
\newcommand{\ta}{\theta}
\newcommand{\f}{\phi}
\newcommand{\vf}{\varphi}
\newcommand{\ka}{\kappa}
\newcommand{\al}{\alpha}
\newcommand{\ga}{\gamma}
\newcommand{\de}{\delta}
\newcommand{\si}{\sigma}
\newcommand{\bomega}{\mbox{\boldmath $\omega$}}
\newcommand{\bsi}{\mbox{\boldmath $\sigma$}}
\newcommand{\bchi}{\mbox{\boldmath $\chi$}}
\newcommand{\bal}{\mbox{\boldmath $\alpha$}}
\newcommand{\bpsi}{\mbox{\boldmath $\psi$}}
\newcommand{\brho}{\mbox{\boldmath $\varrho$}}
\newcommand{\beps}{\mbox{\boldmath $\varepsilon$}}
\newcommand{\bxi}{\mbox{\boldmath $\xi$}}
\newcommand{\bbeta}{\mbox{\boldmath $\beta$}}
\newcommand{\ee}{\end{equation}}
\newcommand{\eea}{\end{eqnarray}}
\newcommand{\be}{\begin{equation}}
\newcommand{\bea}{\begin{eqnarray}}
\newcommand{\ii}{\mbox{i}}
\newcommand{\e}{\mbox{e}}
\newcommand{\pa}{\partial}
\newcommand{\Om}{\Omega}
\newcommand{\vep}{\varepsilon}
\newcommand{\bfph}{{\bf \phi}}
\newcommand{\lm}{\lambda}
\def\theequation{\arabic{equation}} 
\renewcommand{\thefootnote}{\fnsymbol{footnote}}
\newcommand{\re}[1]{(\ref{#1})}
\newcommand{\R}{{\rm I \hspace{-0.52ex} R}}
\newcommand{\N}{{\sf N\hspace*{-1.0ex}\rule{0.15ex}%
{1.3ex}\hspace*{1.0ex}}}
\newcommand{\Q}{{\sf Q\hspace*{-1.1ex}\rule{0.15ex}%
{1.5ex}\hspace*{1.1ex}}}
\newcommand{\C}{{\sf C\hspace*{-0.9ex}\rule{0.15ex}%
{1.3ex}\hspace*{0.9ex}}}
\newcommand{\eins}{1\hspace{-0.56ex}{\rm I}}
\renewcommand{\thefootnote}{\arabic{footnote}}

\title{Self-dual instanton and nonself-dual instanton-antiinstanton 
\\
solutions in $d=4$ Yang-Mills theory}

\author{{\large Eugen Radu}$^{\dagger}$ and {\large D. H. 
Tchrakian}$^{\dagger
\star}$ \\ \\ $^{\dagger}${\small Department of Mathematical Physics, 
National
University of Ireland Maynooth,} \\ {\small Maynooth, Ireland}
\\ $^{\star}${\small School of Theoretical Physics -- DIAS, 10 
Burlington Road,
Dublin 4, Ireland }}

\maketitle

\begin{abstract}
Subjecting the $SU(2)$ Yang--Mills system to azimuthal symmetries in 
both the $x-y$ and the $z-t$ planes results in a residual subsystem
described by a $U(1)$ Higgs like model with two complex scalar 
fields on the quarter plane. The resulting instantons are labeled by 
integers $(m,n_1,n_2)$ with topological charges $q=\frac12\,[1-(-1)^m]n_1n_2$. 
Solutions are constructed numerically for $m=1,2,3$ and a range of
$n_1=n_2=n$. It is found that only the $m=1$ instantons are self-dual, the
$m>1$ configurations describing composite instanton-antiinstanton lumps.
\end{abstract}

\section{Introduction}
After the discovery of the BPST instantons by Belavin {\it et al}~\cite{BPST}, 
which are unit topological charge spherically symmetric solutions to the
$SU(2)$ Yang-Mills (YM) equations, Witten~\cite{W} constructed higher charge
axially symmetric instantons. Multi-instantons subject to no symmetries
were subsequently constructed by 't~Hooft and by Jackiw {\it et al}~\cite{JNR}
for the same system, while the most general instantons for arbitrary 
gauge group $SU(N)$ were classified by Atiyah {\it et al}~\cite{ADHM}. All 
these instantons are solutions to the first order self-duality equations.

Solutions to the second order Euler-Lagrange equations which
are not self-dual, while interesting for their own sake, are of 
great physical importance. The most important such solutions are the zero
topological charge instanton--antiinstantons, whose putative role has 
been studied from the eariest stages of instanton physics, in particular 
in the construction of dilute (or otherwise) instanton gases~\cite{P,CDG}. To 
date however no such exact solutions were constructed, although the 
forces between an instanton and an antiinstanton were considered~\cite{F} 
long ago for approximate field configurations.

The existence of non self-dual instantons has been proved in
\cite{existence},  and for the
$SU(2)$ YM system
in \cite{SS,GB}. In the latter, the residual actions studied 
are one dimensional as a result of the imposition of quadrupole symmetry in 
\cite{SS}, and by applying ``equivariant geometry'' in \cite{GB}. 
However, the proof in \cite{SS} does not cover the existence of non self-dual
instantons with topological charges $\pm 1$. 

In this Letter, we present both self-dual and non self-dual instantons of
the $SU(2)$ YM model, subjected to azimuthal symmetries both in the 
$x-y$ and the $z-t$ planes. The residual system in our case is an axially 
symmetric $U(1)$ Higgs like model in the $\rho-\si$ quarter
plane\footnote{With $\rho^2=x^2+y^2$ and $\si^2=z^2+t^2$, where
$x,y,z$ and $t$ are rectangular cordinates on $E^4$.}, as distinct from
Witten's case~\cite{W} where it is a $U(1)$ Higgs model on the half plane
$(r,t)$. We find finite action solutions labeled by a triple of 
integers $(m,n_1,n_2)$ with topological charges 
$q=\frac12\,[1-(-1)^m]n_1n_2$. We find that only for $m=1$ are the instantons
self-dual, all those labeled by $m\ge 2$ being non self-dual. All our 
solutions are constructed numerically.

\section{The model}
\subsection{Imposition of symmetry and residual action}
The usual $SU(2)$ Yang-Mills (YM) action density in 4 Euclidean 
dimensions
\be
\label{YM}
{\cal L}=\frac{1}{16\pi^2}\,\mbox{Tr}\,{\cal 
F}_{\mu\nu}^2\quad,\quad~{\rm with}~~
{\cal F}_{\mu\nu}=\pa_{\mu}{\cal A}_{\nu}-\pa_{\nu}{\cal A}_{\mu}+
[{\cal A}_{\mu},{\cal A}_{\nu}],
\ee
will be subjected to two successive axial symmetries. The 
normalisation used
in \re{YM} is chosen such that the spherically symmetric 
BPST~\cite{BPST}
instanton has unit topological charge, in the spherically symmetric 
limit of
our Ansatz to be stated below.
We denote the Euclidean four dimensional coordinates as
$x_{\mu}=(x,y;z,t)\equiv(x_{\al};x_i)$, with $\al=1,2$ and $i=3,4$, 
and use the
following parametrisation
\begin{eqnarray}
x_{\al}=r\sin\ta\,\hat x_{\al}\equiv\rho\,\hat x_{\al},
~~~x_{i}=r\cos\ta\,\hat x_{i}\equiv\si\hat x_i\,,
\label{coord}
\end{eqnarray}
where $r^2=|x_{\mu}|^2=|x_{\al}|^2+|x_{i}|^2$, with the unit vectors
appearing in \re{coord} parametrised as
 $\hat x_{\al}=(\cos\vf_1,\sin\vf_1),~~\hat 
x_{i}=(\cos\vf_2,\sin\vf_2)$, 
with $0\le\ta\le\frac{\pi}{2}$ spanning the
quarter plane, and the two azimuthal angles $0\le\vf_1\le 2\pi$
and $0\le\vf_2\le 2\pi$.

The $SU(2)$ YM field ${\cal F}_{\mu\nu}$
will be subjected to two stages of symmetry, in the $x_{\al}$ and 
the
$x_i$ planes, in succession.
The {\bf first stage} of symmetry imposition is of cylindrical 
symmetry in the
$x_{\al}=(x_1,x_2)$ plane. Our {\it cylindrically symmetric} Ansatz 
is
\bea
\label{RR-Aa}
{\cal A}_{\al}&=&\left(\frac{\phi^3+n_1}{\rho}\right)\,\Sigma_{\al\beta}
\hat x_{\beta}+\left(\frac{\phi^1}{\rho}\right)\,(\vep\hat x)_{\al}\,
(\vep n^{(1)})+{\beta}\,\Sigma_{\beta3}
-A_{\rho}^2\,\,\hat x_{\al}n^{(1)}_{\beta}\,\Sigma_{\beta3}\\
\nonumber
&&\qquad\qquad\qquad\qquad
+\left(\frac{\phi^2}{\rho}\right)\,(\vep\hat x)_{\al}\,
(\vep n^{(1)})_{\beta}\,\,\Sigma_{\beta4}
+A_{\rho}^1\,\,\hat x_{\al}\,n^{(1)}_{\beta}\,\Sigma_{\beta4}
-A_{\rho}^3\,\hat x_{\al}\Sigma_{34}~,
\\
{\cal A}_i&=&-A_i^1\, 
n^{(1)}_{\beta}\,\Sigma_{\beta3}+A_i^2\,n^{(1)}_{\beta}\,
\Sigma_{\beta4}-A_i^3\,\Sigma_{34}~,
\label{RR-Ai}
\eea
in terms of the unit vector $n^{(1)}_{\al}=(\cos n_1\vf_1\,,\,\sin n_1\vf_1)$
labeled by the vorticity integer $n_1$, $\vep_{\al\beta}$ being the
Levi-Civita symbol. The spin matrices
$\Sigma_{\mu\nu}=(\Sigma_{\al\beta},\Sigma_{\al i},\Sigma_{ij})$ in
\re{RR-Aa} and \re{RR-Ai} are one or other of the two
chiral representations of $SO(4)$, i.e. they are $SU(2)$ matrices.

Using the notation
\be
\label{curvcyl}
F_{MN}=\pa_M A_N^a-\pa_NA_M^a+\vep^{abc}A_M^bA_N^c~,~~~
D_M\phi^a=\pa_M\phi^a+\vep^{abc}A_M^b\phi^c\,~,
\ee
with $x_M=(x_i,\rho)$ and the internal index $a=1,2,3$, the components of the
curvature ${\cal F}_{\mu\nu}$ is expressed exclusively in terms of 
the gauge covariant quantities \re{curvcyl}, describing an
effective $SO(3)$ YM-Higgs system on the hyperbolic space with 
coordinates $x_M=(x_i,\rho)$.

Imposition of the {\bf second stage} of symmetry is in the $x_i$-plane in
this hyperbolic space, by subjecting the fields
$(A_i^a,A_{\rho}^a)$ and $\phi^a$
appearing in \re{curvcyl}, to axial symmetry in the $x_i=(x_3,x_4)$ plane.
Relabeling the $SO(3)$ internal index as $a=(i',3)$, i.e. 
$i',j',..=1,2$, the axially symmetric Ansatz for the
residual $SO(3)$ gauge connection is
\bea
\nonumber
A_i^{i'}&=&-a_{\si}\,\hat x_i\,(\vep n_{(2)})^{i'}+
\left(\frac{\chi^1}{\si}\right)(\vep\hat x)_i\,n_{(2)}^{i'}~,
\label{ii'}
\\
A_i^3&=&\left(\frac{\chi^2+n_2}{\si}\right)\,(\vep\hat x)_i~,
\label{i3}
\\
\nonumber
A_{\rho}^{i'}&=&-a_{\rho}\,(\vep n_{(2)})^{i'}~,
\label{rhoi'}
\\
\nonumber
A_{\rho}^3&=&0\ ,
\eea
with the unit vector 
$n_{(2)}^{i'}=(\cos n_2\vf_2\,,\,\sin n_2\vf_2)$
labeled by a second vorticity integer $n_2$. Subjecting the $SO(3)$ triplet
effective Higgs field $\phi^a=(\phi^{i'},\phi^3)$ to the
same symmetry, we have $\phi^{i'}=\xi^1\,n_{(2)}^{i'},~~\phi^3=\xi^2\ .$

 In the numerical computation, we implemented a {\bf third stage} of
symmetry by treating the two azimuthal symmetries imposed in the $x-y$ and the
$z-t$ planes on the same footing. Thus we set the two vorticities $n_1$ and
$n_2$, appearing in \re{RR-Aa}-\re{RR-Ai} and \re{i3} respectively, equal,
$n_1=n_2=n$. For the rest of this section however we consider the general case
with the two distinct vorticities. 

As a result of imposing the two stages of symmetry, the YM field is
parametrized by six functions which depend only on $\rho$ and $\si$.
The components of the curvature $(F_{ij}^a,F_{i\rho}^a)$ and the covariant
derivatives $(D_{\rho}\phi^a,D_i\phi^a)$ appearing in
\re{curvcyl} are now expressed exclusively in terms of the $SO(2)$ curvature
\be
\label{finalf}
f_{\rho\si}=\pa_{\rho}a_{\si}-\pa_{\si}a_{\rho}
\ee
and the covariant derivatives
\begin{eqnarray}
{\cal D}_{\rho}\chi^A=\pa_{\rho}\chi^A+
a_{\rho}(\vep\chi)^A\quad,\quad
{\cal D}_{\si}\chi^A=\pa_{\si}\chi^A+
a_{\si}(\vep^\chi)^A,
\label{dchi}
\\
{\cal D}_{\rho}\xi^A=\pa_{\rho}\xi^A+
a_{\rho}(\vep\xi)^A\quad,\quad
{\cal D}_{\si}\xi^A=\pa_{\si}\xi^A+
a_{\si}(\vep\xi)^A,
\nonumber
\end{eqnarray}
where we have used the notation $(\chi^1,\chi^2)=\chi^A$,
$(\xi^1,\xi^2)=\xi^A$, $A=1,2$.
The result is a residual $U(1)$ connection $(a_{\rho},a_{\si})$
interacting covariantly with the scalar
fields $\chi^A$ and $\xi^A$, i.e. an Abelian Higgs like model in the
quarter plane $\rho-\si$, described by a Lagrangean 
\be
\label{redact}
L=\frac14\left[\rho\si\,f_{\rho\si}^2
+\frac{\rho}{\si}\left(|{\cal D}_{\rho}\chi^A|^2
+|{\cal D}_{\si}\chi^A|^2\right)
+\frac{\si}{\rho}\left(|{\cal D}_{\rho}\xi^A|^2
+|{\cal D}_{\si}\xi^A|^2\right)+
\frac{1}{\rho\si}(\vep^{AB}\chi^A\xi^B)^2\right]~.
\ee 
This residual action density is a scalar with
respect to the {\it local} $SO(2)$ indices $A,B$, hence they are
manifestly gauge invariant. It describes a $U(1)$ Higgs like model 
with
{\bf two} effective Higgs fields $\chi^A$ and $\xi^A$, coupled minimally
to the $U(1)$ gauge connection $a_{\mu}=(a_{\rho},a_{\si})$. To remove this
$U(1)$ gauge freedom we impose the usual gauge condition
\be
\label{gc}
\pa_{\mu}\,a_{\mu}\equiv\pa_{\rho}a_{\rho}+\pa_{\si}a_{\si}=0~.
\ee
To state compactly the residual self-duality equations which saturate the
topological lower bound of \re{redact}, we use a new index notation
$x_{\mu}=(\rho,\si)$, not to be confused with the index notation used 
in
\re{YM}. They are expressed in this notation as
\bea
f_{\mu\nu}&=&\frac{1}{\rho\si}\vep_{\mu\nu}\vep^{AB}\chi^A\xi^B~,
\label{f}
\\
{\cal D}_{\mu}\chi^a
&=&{\rho\,}^{-1}\si\,\vep_{\mu\nu}{\cal D}_{\nu}\xi^a\,.
\label{rs}
\eea
One solution of these equations is well known, corresponding to the
spherically symmetric unit charge BPST~\cite{BPST} instanton.
 Expressing the self-duality equations \re{f}-\re{rs} in terms of the
coordinates $(r,\ta)$ rather than $(\rho,\si)$, constraining the function
$a_{\ta}=a_{\ta}(r)$ to be a radial function and assigning the
following values of the remaining fuctions $(a_r,\chi^A,\xi^A)$:
\be
\label{m1n1}
a_r=0,~\chi^1=-\xi^1=\frac{1}{2}a_\theta \sin 2 
\theta,~~
\chi^2=-a_\theta \cos^2 \theta-1,~~\xi^2=-a_\theta \sin^2 \theta-1~,
\ee
these reduce to the single self-duality equation
$d a_{\ta}/dr =a_{\ta}(a_{\ta}+2)/r$
yielding 
\be
\label{bpst}
a_{\ta}=-\frac{2r^2}{r^2+\la^2}
\ee
($\la$ being the arbitrary scale of the unit charge instanton).

Since our numerical constructions will be carried out using the 
coordinates
$(r,\ta)$  we display \re{redact} also as
\bea
\label{redactsph}
L=\frac14\bigg[  r\sin\ta\cos\ta\,f_{r\ta}^2
&+&\frac{r\sin\ta}{\cos\ta}\left(|{\cal D}_{r}\chi^A|^2
+\frac{1}{r^2}|{\cal D}_{\ta}\chi^A|^2\right)
\\
\nonumber
&+&\frac{r\cos\ta}{\sin\ta}\left(|{\cal D}_{r}\xi^A|^2
+\frac{1}{r^2}|{\cal D}_{\ta}\xi^A|^2\right)+
\frac{1}{r\sin\ta\cos\ta}(\vep^{AB}\chi^A\xi^B)^2\bigg]~,
\eea
the total action of these solutions being 
 $S=\int d^4x \sqrt{g}\mathcal {L}=
\int_0^{\infty}\int_0^{\pi/2}dr d\theta L$,
with $\mathcal {L}$ the action density \re{YM}.

 It is worth noting that precisely the same results for the residual gauge
connection and reduced action are found by the alternative approach
of \cite{Forgacs:1980zs}, where the action of the isometries 
$\partial/\partial \varphi_1, ~\partial/\partial \varphi_2$ on the gauge
connection is compensated by suitable gauge transformations. 

\subsection{Boundary conditions}

To obtain regular solutions with finite action density we impose at 
the origin 
($r=0$) the boundary conditions
\be
\label{r0}
a_r=0\quad,\quad a_{\ta}=0\quad,\quad
\chi^A=\left(
\begin{array}{c}
\ 0 \\
-n_2
\end{array}
\right)\quad,\quad
\xi^A=\left(
\begin{array}{c}
\ 0 \\
-n_1
\end{array}
\right)\,~,
\ee
which are requested by the analyticity of the ansatz.
In order to find finite action solutions, we impose at infinity 
\be
\label{rinfty}
a_r=0~,~~a_{\ta}=-2m~,~~\chi^A=
(-1)^{m+1}n_2\,\left(
\begin{array}{c}
\sin 2m\ta \\
\cos 2m\ta
\end{array}
\right)~,~~
\xi^A=-n_1\,\left(
\begin{array}{c}
\sin 2m\ta \\
\cos 2m\ta
\end{array}
\right)\,,
\ee
$m$ being a positive integer.
Similar considerations lead to the following
 boundary conditions on the $\rho$ and $\sigma$ axes:   
\bea
a_r=\frac{1}{n_1}\,\pa_{r}\xi^1\quad,
\quad a_{\ta}=\frac{1}{n_1}\,\pa_{\ta}\xi^1\quad,~~
\chi^1=0\quad,\quad\xi^1=0,~~
\pa_{\ta}\chi^2=0\quad,\quad\xi^2=-n_1~,
\label{th0}
\eea
for  $\theta=0$ and  
\bea
a_r=\frac{1}{n_2}\,\pa_{r}\chi^1\quad,\quad
a_{\ta}=\frac{1}{n_2}\,\pa_{\ta}\chi^1\quad,\quad
\chi^1=0\ \ \quad,\ \ \quad\xi^1=0,~~
\chi^2=-n_2\quad,\quad\pa_{\ta}\xi^2=0\label{thp2}\,~,
\eea
for $\theta=\pi/2$, respectively.

\subsection{Topological charge}
Having determined the values of all the functions on the boundaries 
of
$(r,\ta)$-domain, we proceed to calculate the resulting topological charges.
In our normalisation, the topological charge is defined as
\be
\label{pont}
q=\frac{1}{32\pi^2}\vep_{\mu\nu\rho\si}\int\mbox{Tr}
{\cal F}_{\mu\nu}{\cal F}_{\rho\si}\,d^4x\,,
\ee
which after integration of the azimuthal angles $(\vf_1,\vf_2)$ 
reduces to
\be
\label{topch1}
q=\frac12\int\left(\vep_{AB}\chi^A\xi^B\,f_{\rho\si}+
{\cal D}_{[\rho}\chi^A\,{\cal D}_{\si]}\xi^A\right)\,d\rho\,d\si\,,
\ee
which in the index notation used in \re{f}-\re{rs} can be expressed 
compactly
as
\bea
q&=&\frac12\vep_{\mu\nu}\,\int\left(\frac12\,\vep_{AB}\chi^A\xi^B\,f_{\mu\nu}+
{\cal D}_{\mu}\chi^A\,{\cal 
D}_{\nu}\xi^A\right)\,d^2x\label{topch2}\\
&=&\frac14\int\vep_{\mu\nu}\,\pa_{\mu}(\chi^A{\cal D}_{\nu}\xi^A-
\xi^A{\cal D}_{\nu}\chi^A)\,d^2x\,.\label{totdiv}
\eea
The integration in \re{topch2} is carried out over the 2 dimensional
space $x_{\mu}=(x_{\rho},x_{\si})$. 
As expected this is a total
divergence expressed by \re{totdiv}.

\newpage
\setlength{\unitlength}{1cm}

\begin{picture}(7,5.5)
\centering
\put(2,-1.0){\epsfig{file=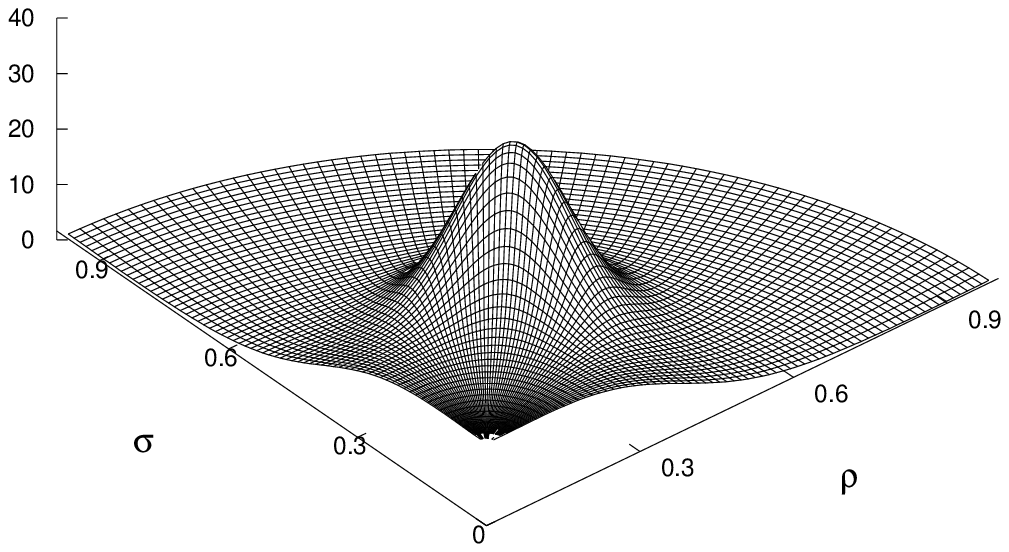,width=14cm}}
\end{picture}
\\
\\
\\
{\small {\bf Figure 1.} 
The action density $\mathcal{L}$ is shown for the $m=1,~n=4$ 
self-dual
instanton (with $\rho=r\sin \theta$ and $\sigma=r \cos \theta$).}
\\
\\
Using Stokes' theorem, the two dimensional integral of \re{totdiv} 
reduces
to the one dimensional line integral
\be
\label{stokes}
q=\frac14\int
\chi^A\stackrel{\leftrightarrow}{\cal D}_{\mu}\xi^A\,ds_{\mu}\,,
\ee
the line integral being taken around the loop
\be
\label{loop}
q=\frac14\left(\int_{0}^{\infty}
\left(\chi^A\stackrel{\leftrightarrow}{\cal D}_{\si}\xi^A\right)
\bigg|_{\rho=0}\,d\si+
\int_{0}^{\frac{\pi}{2}}
\left(\chi^A\stackrel{\leftrightarrow}{\cal D}_{\ta}\xi^A\right)
\bigg|_{r=\infty}\,d\ta+
\int_{\infty}^{0}
\left(\chi^A\stackrel{\leftrightarrow}{\cal D}_{\rho}\xi^A\right)
\bigg|_{\si=0}\,d\rho\right)\,.
\ee
The $\ta$ integral over the large quarter circle vanishes since
${\cal D}_{\ta}\chi^A$ and ${\cal D}_{\ta}\xi^A$ both vanish as 
$r\to\infty$,
so the the only contributions come from the $\si$ and $\rho$ 
integrations.
These are immediately evaluated by reading off the appropriate values 
of
$\chi^A$ and $\xi^A$ from \re{th0}-\re{thp2}. The result is
\be
\label{q}
q=\frac12\,[1-(-1)^m]n_1n_2\,,
\ee
such that only for odd $m$ is the Pontryagin charge nonzero and is 
then
always equal to $n_1n_2$. This is consistent with the description of
instanton-antiinstanton chains. Obviously, if zero topological
charge solutions exist, they must be non self-dual, as also must the 
higher odd
$m\ge 3$ solutions solutions with nonzero charge since their actions
are likely to grow with $m$ while the (nontrivial) lower bound stays 
the
same at $n_1n_2$. It is likely that the odd $m=1$ solution does 
saturate
this bound and hence is self-dual.  We have verified all these 
features
in our numerical constructions below.

\section{Numerical results}
Subject to the above boundary conditions \re{th0}-\re{thp2}
we solve numerically the
set of six coupled non-linear elliptic partial differential equations.
We employed a compactified radial coordinate 
\newpage
\setlength{\unitlength}{1cm}

\begin{picture}(18,7)
\centering
\put(2,0.0){\epsfig{file=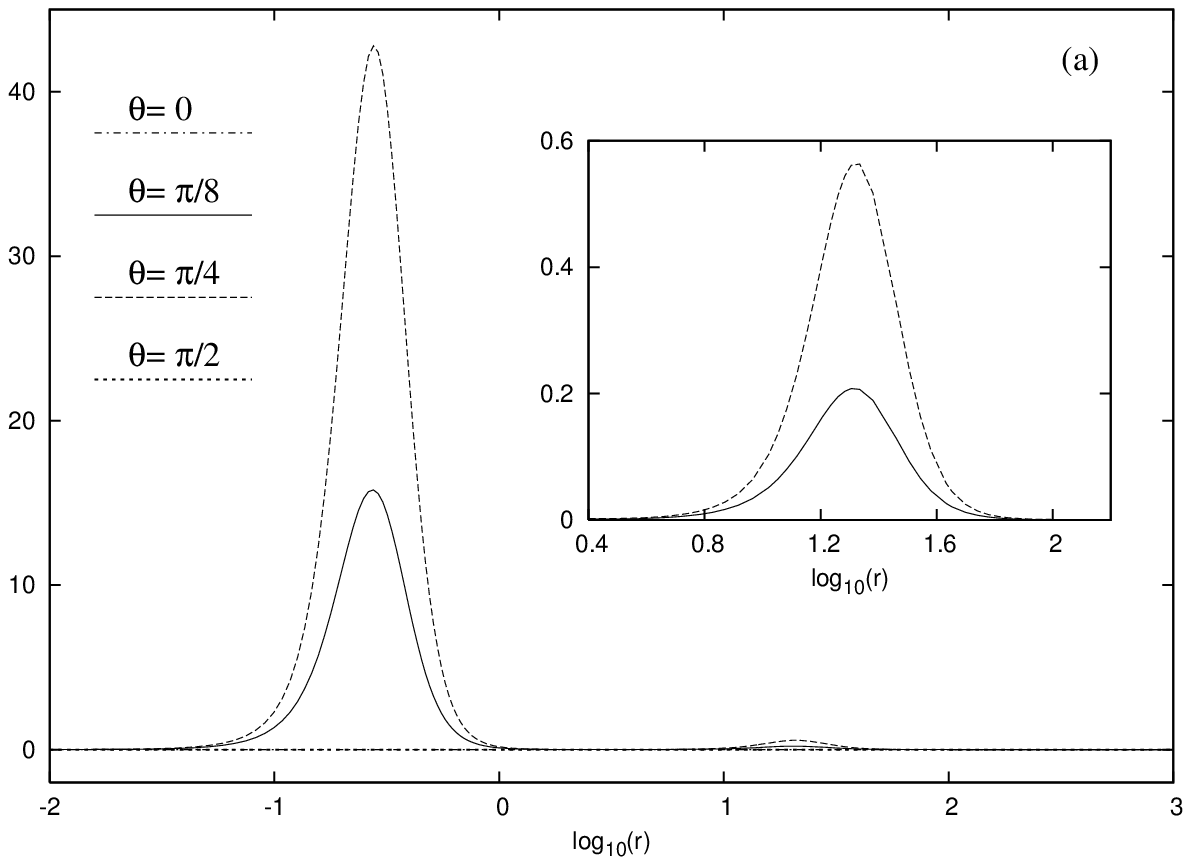,width=11cm}}
\end{picture}
\begin{picture}(19,8.)
\centering
\put(2.6,0.0){\epsfig{file=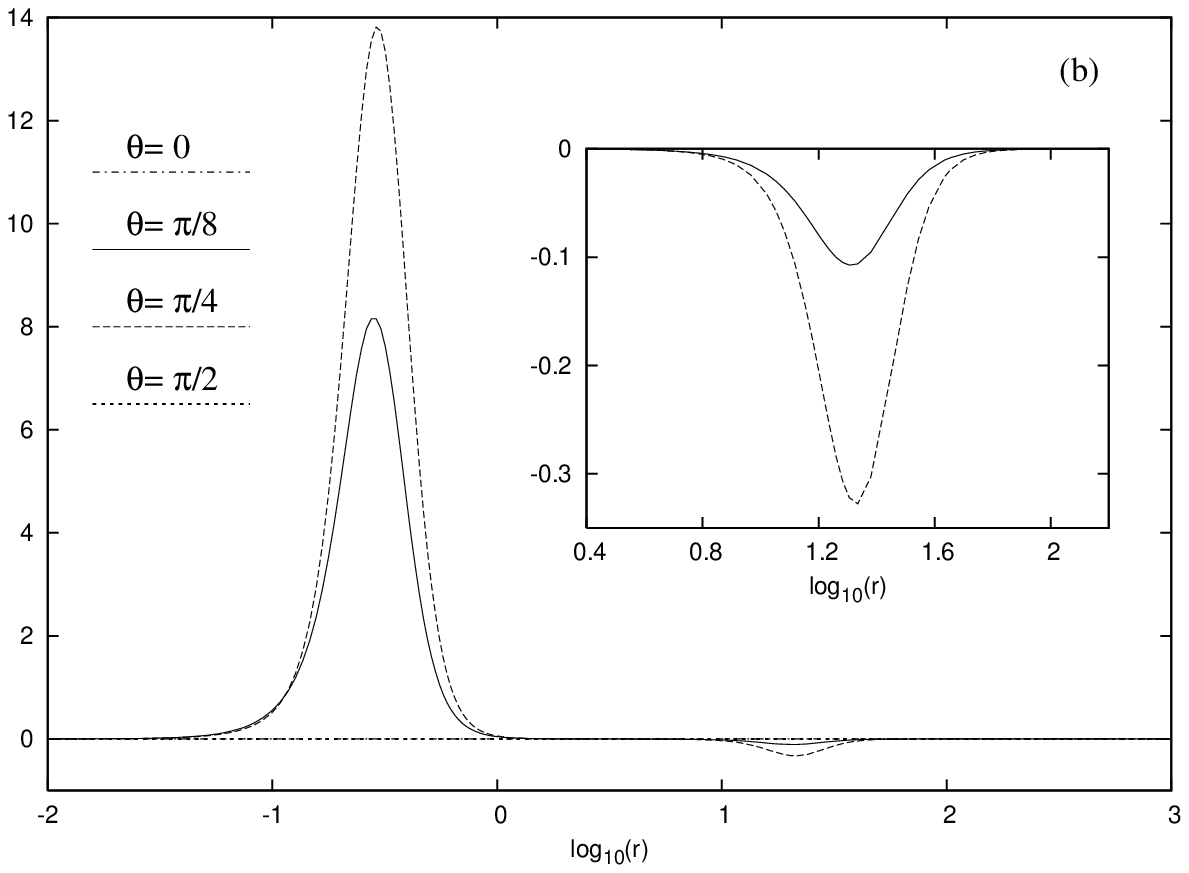,width=11cm}}
\end{picture}
\\
\\
{\small {\bf Figure 2.} The effective
Lagrangean (a) and the topological charge density (b)
as given by (\ref{redactsph}) and (\ref{topch2}) respectively,
are plotted for the $m=2$, $n=2$ instanton-antiinstanton solution.}
\\
\\
$x=r/(1+r)$ in our 
computations,
the equations being discretised on a nonequidistant grid in $r$ and 
$\theta$ with typical grid size $130\times 60$.
The numerical calculations were performed with the software 
package FIDISOL, based on the Newton-Raphson method \cite{FIDISOL}.

To simplify the general picture we set $n_1=n_2=n$. We have studied 
solutions for $m=1$ with  $1\leq n\leq 5$ and for $m=2$ with $1\leq n\leq 3$.
Our preliminary numerical results indicate  that there exist 
also solutions with $m=3$, and hence likely for all $(m,n)$.

Since the $m=1$ solutions are special, we start by discussing their
properties. It turns out that all solutions
with $m=1$ are self-dual.   This was verified by checking numerically
that the solutions of the second order equations satisfy the first order
self-duality equations  \re{f}-\re{rs}.  
\newpage
\setlength{\unitlength}{1cm}

\begin{picture}(18,7)
\centering
\put(2,0.0){\epsfig{file=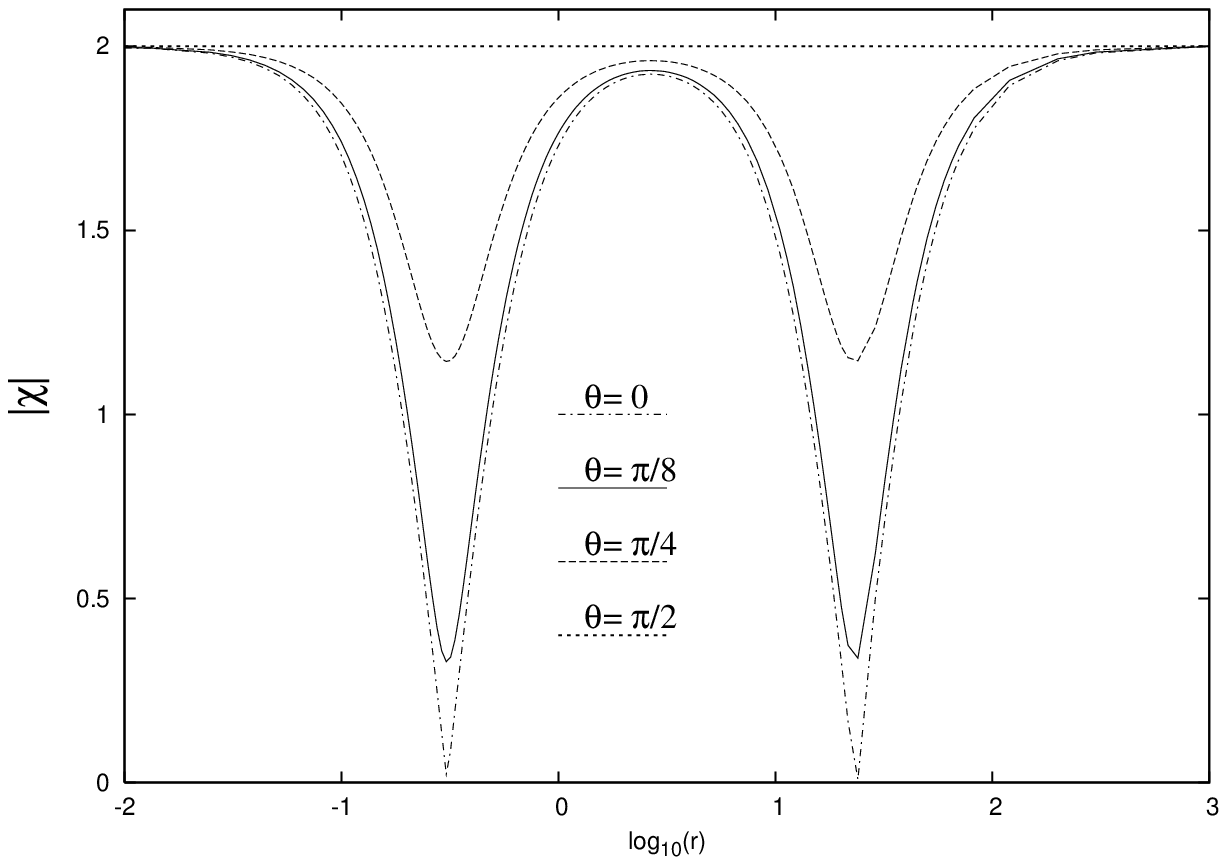,width=11cm}}
\end{picture}
\begin{picture}(19,8.)
\centering
\put(2.6,0.0){\epsfig{file=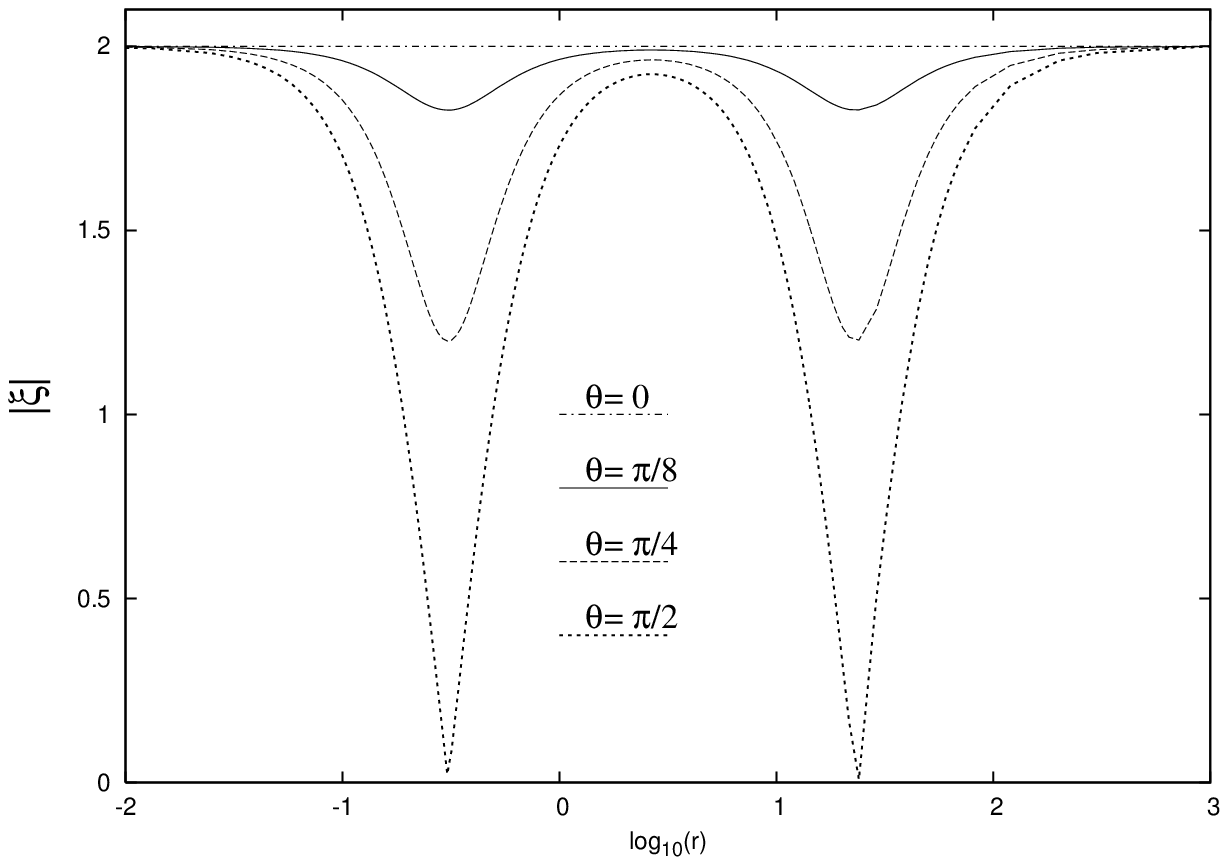,width=11cm}}
\end{picture}
\\
\\
{\small {\bf Figure 3.} The  moduli $|\chi|$ and $|\xi|$ of the effective Higgs
fields of the $m=2,~n=2$ solution are shown as a function of 
$r$ for several angles.}
\\
\\
Moreover the $(m=1,n=1)$ solution is spherically symmetric,
corresponding to the unit
charge BPST instanton (\ref{m1n1}), (\ref{bpst}). 
The $m=1$ solutions with $n\geq 2$ are axially
symmetric and their gauge potentials $a_r,~a_\theta,~\chi^A,~\xi^A$ have
nontrivial $\theta$-dependence. A three dimensional plot of the 
action density for the $m=1,~n=4$ self-dual instanton is presented in Figure 1.

The $m>1$ configurations satisfy only the second order 
Euler-lagrange field equations and are not self-dual. 
The function $a_\theta$ does not exhibit a strong angular 
dependence, while $\chi_1$ and $\xi_1$ have rather similar shapes.
The package FIDISOL provides an error estimate for each unknown function.
The typical numerical error is estimated to  be 
on the order of $10^{-3}$, except for the $n=1$ solutions which are somehow 
special.
\newpage
\setlength{\unitlength}{1cm}

\begin{picture}(18,7)
\centering
\put(2,0.0){\epsfig{file=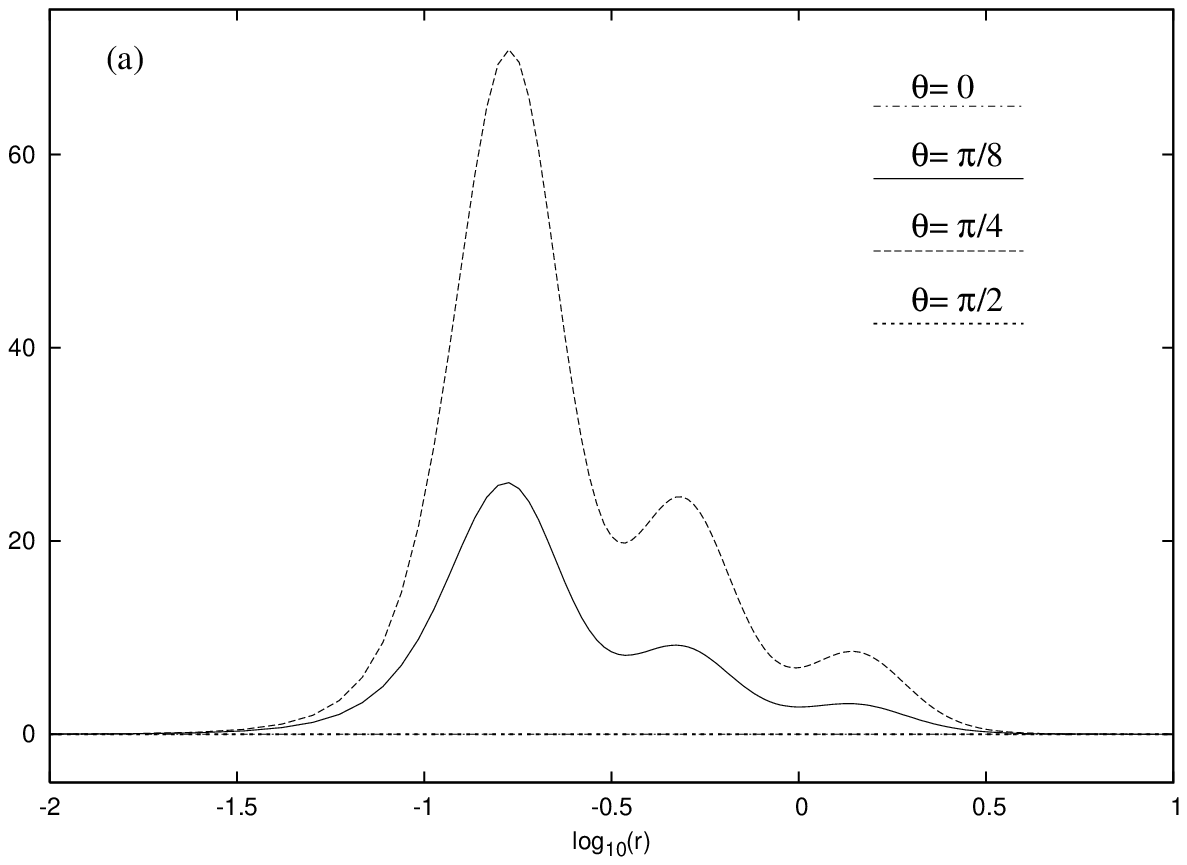,width=11cm}}
\end{picture}
\begin{picture}(19,8.)
\centering 
\put(2.6,0.0){\epsfig{file=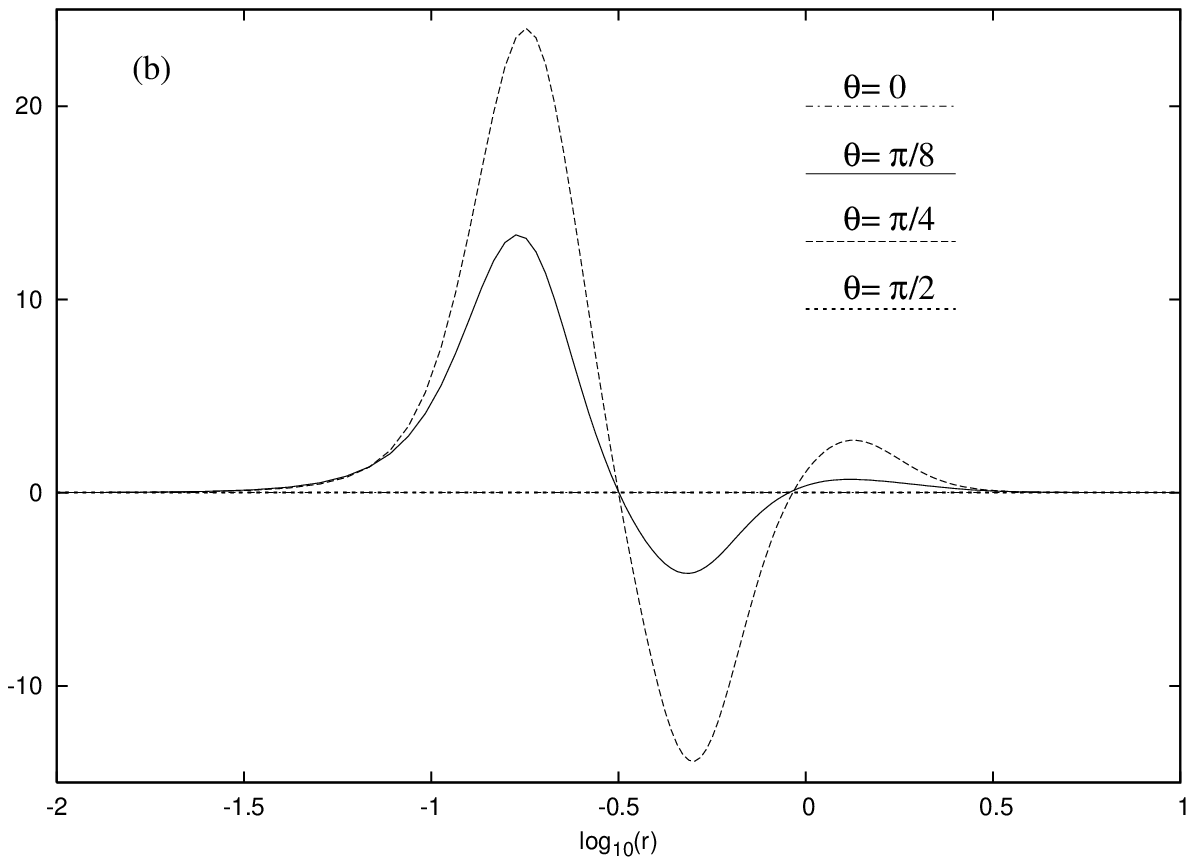,width=11cm}}
\end{picture}
\\
\\
{\small {\bf Figure 4.} Same as Figure 2 for the $m=3$, $n=2$ 
solution.}
 \\
 \\
In this case, although the numerical iteration still
converges, the error is larger. This seems to originate in the
behaviour of the function $a_r$ which has a rather
small $\theta-$dependence (as opposed
to the $n>1$ configurations) and takes very small values.
Its maximal error is around $4\%$ and comes from the vicinity of the 
origin. Thus our numerical results in this case are less conclusive,
the existence and the properties of the $m>1$ solutions with
$n=1$ requiring further work.

A general feature of the solutions we found
is that the action density $\mathcal{L}$ (or the Lagrangean $L$) 
posseses $m$ maxima on the $\theta=\pi/4$ axis.
Thus, in all cases studied it is possible to distinguish 
$m$ individual concentrations of action, the relative distance 
between them
being fixed by the location of the maximum values of the Lagrangean 
$L$
(see Figures 2a and 4a).

\newpage
\setlength{\unitlength}{1cm}

\begin{picture}(18,7)
\centering
\put(2,0.0){\epsfig{file=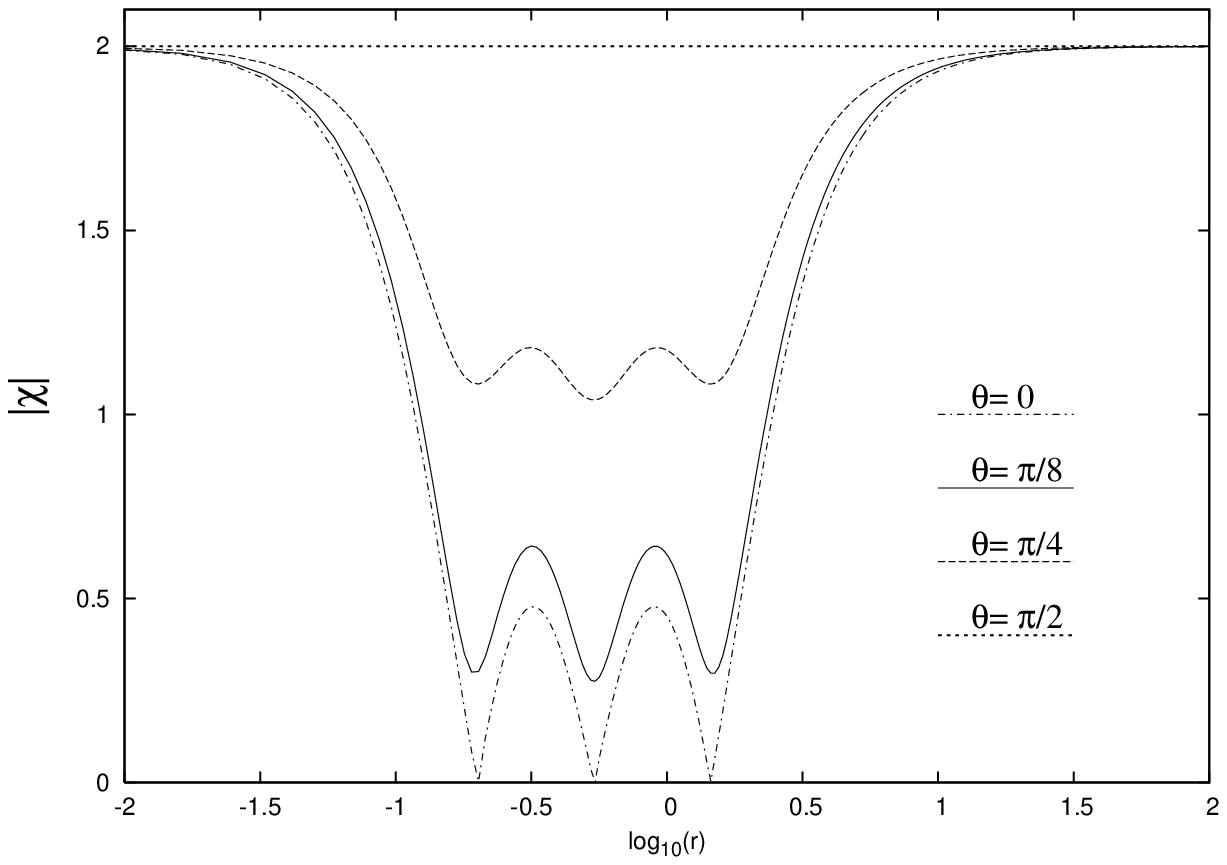,width=11cm}}
\end{picture}
\begin{picture}(19,8.)
\centering
\put(2.6,0.0){\epsfig{file=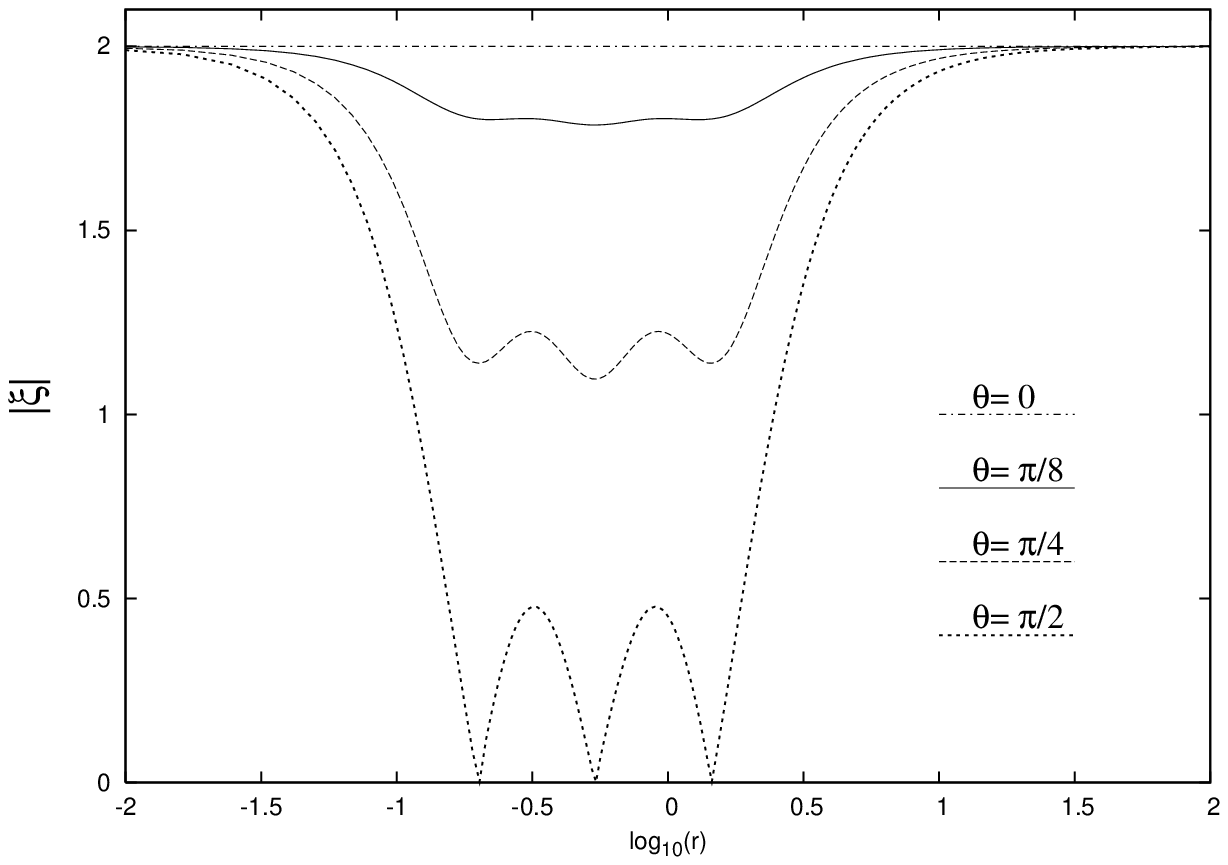,width=11cm}}
\end{picture}
\\
\\
{\small {\bf Figure 5.} Same as Figure 3 for the $m=3$, $n=2$
solution.}
\\
\\
Increasing the value of $n$ causes these maxima to become sharper but 
does not significantly 
affect the distances between the locations of these
pseudoparticles. It is also interesting to note 
that, as shown in 
Figures 3 and 5 for the $m=2,~3$ configurations with $n=2$, the moduli of the
effective Higgs fields $|\chi|=(\chi^A\chi^A)^{1/2}$ and
$|\xi|=(\xi^A\xi^A)^{1/2}$ possess $m$ nodes on the $\rho$ and 
$\sigma$ axis, respectively. The positions of these nodes coincides, within
the numerical accuracy, with the locations of individual pseudoparticles.

The topological charge density, namely the integrand of 
(\ref{topch2}),
also presents $m$ local extrema on the $\theta=\pi/4$ axis, whose  
locations
always coincide with the action density extrema.
However, as shown in Figures 2b and 4b, the signs of the charges
alternate between the locations of the 
successive lumps.

Thus the $m=1$ configurations describe self-dual instanton 
solutions,
$m=2$ corresponds to an instanton-antiinstanton pair,
while the $m>2$ solutions are instanton-antiinstanton bound states
composed of several pseudoparticles.

The numerical results indicate that the action $S(m,n)$ of a composite
$(m,n)-$solution (with $m>1$) is smaller than the action $S_0(m,n)=mn^2$, of
$m$  single infinitely separated self-dual instantons each with vorticity $n$.
We found e.g. $S(2,2)=7.64$, $S(2,3)=15.96$ while $S(3,2)=11.38$.

Also, we verified numerically by integrating (\ref{topch2}) that,
within the numerical accuracy,
the solutions with $m=3$ carry topological charge $n^2$, while the 
topological
charge of $m=2$ solutions vanishes for all values of $n$.

\section{Summary and discussion}
To summarise,
we have constructed instantons of the four dimensional $SU(2)$ YM 
system by numerically integrating the second order Euler--Lagrange 
equations of the residual two dimensional subsystem resulting from the 
imposition of azimuthal symmetries in 
both the $x-y$ and the $z-t$ planes. 
The residual system is a $U(1)$ Higgs like model
featuring  two complex scalar functions. The instantons are labeled by a
triple of integers $(m,n_1,n_2)$ and have topological charges 
$q=\frac12\,[1-(-1)^m]n_1n_2$. 
Concrete constructions were presented for the cases $m=1,2,3$ and several
values of $n_1=n_2=n$,  although nontrivial solutions are likely to 
exist for any positive $(m,n_1,n_2)$.

Clearly, all solutions with even $m$ carry vanishing topological charge,
while all those with odd $m$ carry charge $n_1\,n_2=n^2$, in our case.
Thus our solutions
describe instanton-antiinstanton lumps, rather analogous to the
monopole-antimonopole chains of the Yang-Mills-Higgs model \cite{KKS}.

With the exception of the instantons with $m=1$, all the rest (with $m\ge 2$)
do not saturate the topological bound and are only solutions to the second
order field equations. They are non self-dual solutions with vorticity $n$,
whose actions are always smaller than the action of $m$ infinitely separated
charge-$n^2$, $m=1$ self-dual
instantons. This is supported by our numerical results. Instantons with $m=1$
by contrast, do saturate the topological lower bound and satisfy the first
order self-duality equations. Of these, the $n=1$ solution is the charge-$1$
spherically symmetric BPST instanton, while $m=1$ instantons of charges
$n^2$ are not spherically symmetric. These features are also borne out by our
numerical results.
 
The instantons and antiinstantons  are located in alternating
order on the $\rho=\sigma $ symmetry axis, at the nodes of the moduli
$|\chi^A|$ and $|\xi^A|$ of effective Higgs fields. A detailed study of these
solutions will be presented in a future study. 

Concerning the relation of our results with those of \cite{SS,GB} proving
existence of non self-dual instantons, it may be interesting to speculate as
follows: In the latter~\cite{SS,GB}, it is stated that the proof of
existence for non self-dual instantons of charge $|q|=1$ is absent. In our
case, while the self-dual solution with $m=1,~n=1$ is evaluated in closed
form, the $m\ge 2,~n=1$ non self-dual instantons of charge $q=n^2=1$ are
the only solutions for which the numerical process had some difficulties,
discussed in Section 3. This is the class of non self-dual instantons for
which the existence proofs of $m\ge 2,~n=1$ are absent.

On a physical level, we hope that our solutions will prove useful both in
the sense that they are labeled by integers $(m,n)$ and describe higher
charged instantons, and especially since they present exact non self-dual
solutions that can be useful in the construction of instanton gases and
liquids. Hopefully, these numerical results will be of help in constructing
the solutions analytically.

\bigskip
\noindent
{\bf\large Acknowledgements} \\
We are indepted to A. Chakrabarti, P. Forg{\'a}cs,
D. Maison and W. Nahm for very helpful discussions.
This work is carried out
in the framework of Enterprise--Ireland Basic Science Research 
Project
SC/2003/390 of Enterprise-Ireland.

\begin{small}

\end{small}

\end{document}